%% file: my808_astroph.tex
\documentstyle{mn}

\newif\ifAMStwofonts

\input psfig2
\def\hh{$^h$}
\def\mm{$^m$}
\def\ss{$^s$}
\def\deg{\degr}
\def\asec{\arcsec}
\def\amin{\arcmin}
\def\sqamin{\hbox{\sq\amin}}
\def\etal{\rm et al.}
\def\p{$\pm$}
\def\msol{$M_{\sun}$}
\def\rsol{$R_{\sun}$}
\def\teff{$T_{\rm eff}$}
\def\sol{$_{\sun}$}
\def\martin{Mart\'{\i}n}

\title[Searching for Weather in Brown Dwarfs]
      {Searching for Weather in Brown Dwarfs}
\author[C. G. Tinney \& A. J. Tolley]
       {C. G. Tinney$^{1}$\thanks{E-mail: cgt@aaoepp.aao.gov.au} \& A. J. Tolley$^{1,2}$ \\
        $^1$Anglo-Australian Observatory, PO Box 296, Epping. N.S.W. 1710. Australia.\\
        $^2$Jesus College, Oxford University\\
       }
\date{Accepted 13 November 1998.
      Received 7 November 1998;
      in original form 14 September 1998}

\pagerange{\pageref{firstpage}--\pageref{lastpage}}
\pubyear{1999}

\begin{document}

\maketitle

\label{firstpage}

\begin{abstract}
We have used an innovative tuneable filter technique to carry out a search for
the variability signatures of meterological processes in the atmospheres of two
nearby brown dwarfs. We find no evidence for variability in the L-type
brown dwarf DENIS-P\,J1228-1547 in an observations spanning
approximately half a rotation
period (3 hours), and evidence for variability in the M-type
brown dwarf LP\,944-20 in multiple observations spanning approximately one third of
a rotation period (1.5 hours).
\end{abstract}


\section{Introduction}

The discovery of large numbers of brown dwarfs in recent years (q.v. Rebolo \etal\ 1998; 
Zapatero Osorio \etal\ 1999; Tinney 1998b)
has meant that this field can finally
move beyond the ``Guinness Book of World Records'' phase, and into a period where real
understanding of brown dwarf properties is possible. Several major results have
emerged from this increased understanding. The first of these is the
fundamental importance of dust formation in the atmospheres of brown
dwarfs (Allard \etal\ 1997; Tinney \etal\ 1998; Burrows \& Sharp 1998).
A few examples include: the dramatic change seen in the optical spectra of
cool dwarf atmospheres as  effective temperature
(\teff) drops below about 2000K (eg. Kirkpatrick \etal\ 1998),
which is due to the depletion of plasma-phase TiO and VO as
dust containing Ti and V condenses; and that 
clouds of yet another -- presently unidentified -- set of condensates lying above the
photosphere, are thought to be reponsible for the unusually steep optical spectrum
of the brown dwarf Gl\,229B (Schultz \etal\ 1998; Oppenheimer \etal\ 1998; Burrows \& Sharp 1998).

A second fundamental property is that a large number of brown dwarfs are
seen to be rapidly rotating (Mart\'{\i}n \etal\ 1997;
Tinney \& Reid 1998), with
rotational velocities of $v\sin i \approx 20-40$\,km/s being common. At the typical
radii of brown dwarfs (0.1\,\rsol, Burrows \etal\ 1989) this corresponds to rotation 
periods of 6-3 hours for brown dwarfs viewed equatorially. Similar rotation periods 
have been observed in a few of the very lowest mass stars (Mart\'{\i}n \etal\ 1996). 
The combination of rapid rotation with
cloud deck formation, logically leads to the prospect that rotationally
driven meteorology
may play as important a role in brown dwarfs, as it does in 
giant planets. Already observations suggest that circulation
from lower to higher regions of the atmosphere of the Gl\,229B
may take place, resulting in the mixing of CO to the brown dwarf's
surface (Noll, Geballe \& Marley 1997; Oppenheimer \etal\ 1998).

Since, however, we cannot obtain resolved images for brown dwarfs (as we can
for the Solar system giant planets) detecting the signatures of meteorological
effects must be made via time variability. Using Jupiter as a model, we would
expect to observe variability; (1) on time scales similar 
to the rotation period of a brown dwarf (several hours), as storm features appear
from beyond the limb, transit the disc, and vanish again; and, (2) on
the timescales over which meteorological structures form and
dissipate (weeks to months to centuries in the case of Jupiter). 
The detection of these effects will be complicated
by the fact that low-mass stars are known to show variability due to the
presence of spots and flares associated with stellar activity. 
Flares are observed as extremely high temperature events, producing
significant ultra-violet radiation and strong line emission. The
observations we describe below, are extremely insensitive to such
effects. Star spots, however, are observed as regions of depressed
effective temperature passing over the visible disc. This means they
will have a similar timescale to the passing of meterological features
across the disc.

Most of the rapidly rotating cool dwarfs discovered to date
show extremely small levels of chromospheric activity (Tinney \& Reid 1998; 
Basri \& Marcy 1995). It is therefore likely that magnetically induced spots 
will be weak, or absent in these objects. However, even if spots are present,
they will not display the simple ``dark patch'' signatures seen
in hotter stars, because their lower temperatures will
signficantly enhance dust condensation. Moreover, any spots that are present 
will almost certainly interact with the ``unspotted'' level of
cloud formation, and thereby contribute to the overall weather
patterns in brown dwarfs. As our current understanding of
both magnetically driven spot activity, and rotationally driven
storm activity in brown dwarfs is meager indeed, observations of
either or both are interesting -- even if at present we cannot
disentangle their effects. 

We have carried out a first search for
variability of a spectroscopic signature in brown dwarfs, which is sensitive to changes 
in \teff\ and/or the extent of molecular depletion, by using an new charge-shuffling and 
frequency switching technique with the Anglo-Australian Observatory's 
Taurus Tuneable Filter (TTF, Bland-Hawthorn \& Jones 1998a,b).
This technique extends that used by Deutsch, Margon \& Bland-Hawthorn (1998) 
to study time variability at a single wavelength, by the use of
multiple wavelengths. In particular, we have focussed on the
two wavelength ranges shown in Fig. \ref{spectra}. The blue band (B1) lies on
a strong TiO absorption feature in M-dwarfs at $\approx 2000$K.
The strength of this band decreases with decreasing temperature as condensates
containing Ti deplete the atmosphere of plasma-phase TiO. Measuring the
colour difference between this band and a redder band (B2 in Fig. \ref{spectra})
as a function of time should therefore be sensitive to the passage of regions
of different \teff\ (and/or regions of increased or decreased
dust condensation) across the visible disc of a brown dwarf.

\section{Observations}

Observations were carried out on the nights of 1998 February 25 and 26, and 1998
August 20 and 21 (UT), using the 3.9m Anglo-Australian Telescope with TTF
and the MITLL2 2K$\times$4K 15$\mu$m pixel CCD. 
TTF comprises two high finesse, small 
spacing (2=12$\mu$m) etalons designed to deliver superior throughput,
narrow-band imaging in the wavelength range 0.37-1.0$\mu$m (Bland-Hawthorn \& Jones 1998a,b). Its operation
has been integrated with the charge shuffling facilities of the AAO-1 CCD controllers
to permit multiple exposures with rapid ($\sim$1s) frequency switching. For these
runs, the instrument was configured to operate through the TTF I$_6$ blocking filter ($\lambda$=8380-8750\AA)
in two passbands centered near 8570\AA\ (B1) and
8725\AA\ (B2), with a full width at half-maximum (FWHM) spectral resolution of 30\AA. The exact 
central wavelengths measured on each of the two runs are given in Table \ref{log}.
The observations carried out in 1998 February were thought to be photometric
(though no flux standards were observed as our program is entirely differential), but
those obtained in 1998 August were affected by cloud and fog, which severely
limited the amount of data which could be obtained.

\begin{table*}
  \center
  \caption{Observing Log}
  \label{log}
  \begin{tabular}{lccccccccc}
Object & \multicolumn{2}{c}{$\alpha$,$\delta$$^a$} & I$^a$ 
                                                 & UT Date$^b$             &Exposure            & B1$^c$ & B2$^c$ & $\Delta\lambda^d$\\
       & \multicolumn{2}{c}{(J2000.0)}   &       &                         &(s)                 & (\AA)  & (\AA)  &      (\AA) \\[4pt]
DENIS-P\,J1228-1547 & 12:28:13.8&$-$15:47:11&18.2& 1998 Feb 25, 13:54:39.9 &6$\times$30$\times$60& 8570.6 & 8725.4 & 30\\ 
LP\,944-20          & 03:39:34.6&$-$35:25:51&14.2& 1998 Feb 26, 09:59:01.0 &4$\times$30$\times$60& 8570.6 & 8725.4 & 30\\ 
LP\,944-20          & 03:39:34.6&$-$35:25:51&"   & 1998 Aug 20, 16:28:51.1 &4$\times$30$\times$60& 8571.0 & 8723.1 & 30\\ 
LP\,944-20          & 03:39:34.6&$-$35:25:51&"   & 1998 Aug 21, 14:56:39.2 &4$\times$30$\times$60& 8571.0 & 8723.1 & 30\\[4pt]
   \end{tabular}
\raggedright
\noindent
\vskip 10pt
$a$ -- Positions and magnitudes due to Delfosse \etal\ 1997 and Tinney 1996. These references also contain finding charts.\\
$b$ -- This is the time at which the shutter was opened for the first
       exposure taken on this object on this date.\\
$c$ -- Central wavelengths for B1 and B2 were slightly different between
       February and August observing runs.\\
$d$ -- Bandwidth as FWHM, which are the same for both bands. The band-passes are
       extremely well approximated as Lorentzian.
\end{table*}

\begin{figure}
  \centerline{
         \psfig{file=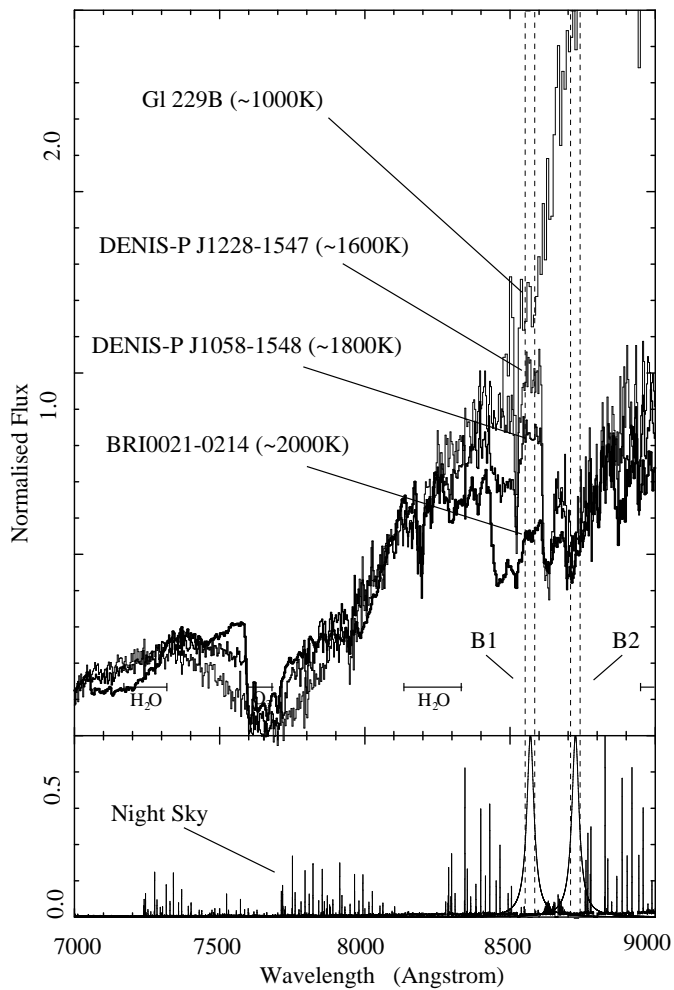,width=3.5in}
            }
\caption{Location of the B1 and B2 bands used in our experiment. The lower panel
shows the B1 and B2 band-passes adopted, superimposed on a night sky spectrum with
the locations of terrestrial absorption marked. The upper panel highlights the variation
in the B1-B2 colour as a function of \teff, via a sequence
of three cool dwarfs, which have been normalized to have the same flux in the
I band. The spectra and approximate \teff\ shown are due to 
Delfosse \etal\ 1997, \martin\ \etal\ 1997, Allard \etal\ 1996, Tinney \etal\ 1998
\& Oppenheimer \etal\ 1998.}
\label{spectra}
\end{figure}

Figure \ref{spectra} shows where the B1 and B2 band-passes fall in relation to the night-sky spectrum,
and the spectra of a cool dwarf sequence. The wavelengths chosen avoid bright
night-sky emission, with B1 lying at a wavelength showing strong TiO absorption in \teff$\approx2000$K
M-dwarf spectra, which is weak or absent in cooler spectra. The B2 band, on the other hand, 
lies in a wavelength range almost unaffected by changes in \teff\ down to the coolest
L-dwarf ($\approx1600$K). At $\approx 1000$K dicrete molecular absorption seems to be
totally absent in both bands, but it is likely that broad H$_2$O absorption is present
(M.Marley, priv.comm.) in both, and that the band fluxes are determined by a combination
of small particle dust opacity (Burrows \& Sharp 1998) and H$_2$O absorption.

\begin{figure}
 \centerline{
%
%
         \psfig{file=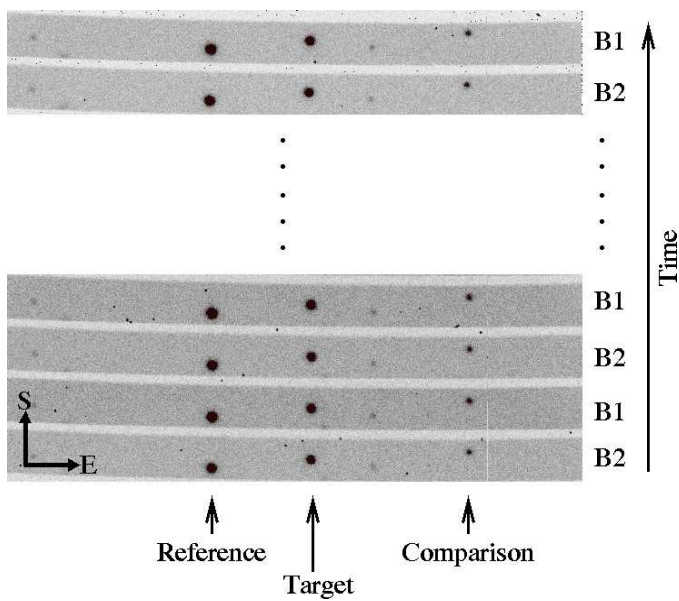,width=3.5in}
            }
\caption{TTF shuffle data for LP\,944-20 from 1998 August 21 (UT), showing
the nature of the time series images acquired, and the
reference, comparison and target stars used (cf. section \ref{rtc}). The curved
field edges are due to the astrometric distortion present at the edge of the Taurus\,2 field.}
 \label{picture}
\end{figure}

Observations were performed through a specially constructed ``slot'' mask, which gave a
28\asec$\times$6.2\amin\ (or 78$\times$1000 pixel) field of view. Because this field is considerably
smaller than our 2K$\times$4K CCD, we are able to use charge shuffling to acquire up to 30 
independent observations of the ``slot'' field before CCD readout is required. For these 
observations (cf. Fig. \ref{picture}) a single exposure
consists of a shuffle sequence as follows: (1) a 60s exposure in B1;
(2) 1s of dead time while TTF is re-configured and charge
is shuffled; (3) a 60s exposure in B2; and (4) another 1s of dead time. This sequence is
repeated 15 times, after the detector is read out at a gain of 1.1e$^-$/adu
and a read noise of 2e$^-$. Each 30 minute observation therefore produces a 1000$\times$2740 pixel
image, comprising fifteen consecutive 60s observations in B1 interleaved with fifteen consecutive 60s observations in B2.

The objects observed are listed in Table \ref{log}. DENIS-P\,J1228-1547 is an L-type (Kirkpatrick \etal\ 1998) 
brown dwarf with mass less than 0.065\,\msol, and age less than 1\,Gyr 
(Tinney, Delfosse \& Forveille 1997; Mart\'{\i}n \etal\ 1997).  Mart\'{\i}n \etal\ (1997) 
report a rotation velocity of $v\sin i \approx 20$\,km/s for DENIS-P\,J1228-1547.
LP\,944-20 (also known
as BRI\,0337-3535) has a mass of 0.060$\pm$0.004\,\msol\ and an age in the range 475-650\,Myr
(Tinney 1998a). Tinney \& Reid (1998) report a rotation velocity of $v\sin i = 28$$\pm$2\,km/s for LP\,944-20.

\begin{figure}
 \centerline{
%
%
         \psfig{file=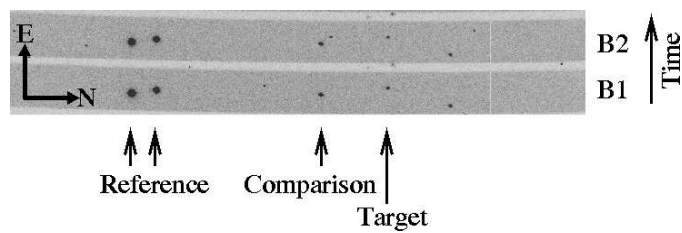,width=3.5in}
            }
\caption{TTF shuffle data for DENIS-P\,J1228-1547 from 1998 February 25 (UT), 
showing the reference, comparison and target stars used (cf. section \ref{rtc}).
Note that the order in which the B1 and B2 bands were observed is reversed from that used in
the August run.}
 \label{pictureb}
\end{figure}

\section{Analysis}

\subsection{Image processing}

Each of the resulting images was bias subtracted using a
20 pixel overscan region. Domeflat observations were taken 
in exactly the same manner as target observations. In order to avoid fluctuations
in the domeflat illumination during the course of a shuffle sequence, these
domeflats were broken up into thirty 1000$\times$90 pixel sub-segments. Each
of the 15 ``copies'' of the B1 and B2 sub-segments were then averaged, and
replicated back into a  full 1000$\times$2740 pixel image. The flattened
data images displayed $\approx$5\% residual variations in sky background,
indicating that in the future, more care will be required in obtaining
flat-fields for this type of data. However, since this experiment aims to
perform {\em differential} photometry, this is not considered a serious
flaw of the current data set. In order to ensure that un-illuminated parts
of the shuffle image are ignored in subsequent photometry, the flat fields
were also used to contruct a mask image, which flagged all un-illuminated pixels
as bad. 

\subsection{Calibration}

A CuAr arc spectrum, acquired
by stepping through a large range in plate spacing, was used to derive wavelength
calibrations, and to confirm the 30\AA\ (FWHM) resolution. Observations were also made 
on February 25 of the standard
star CD-32\,9927 (Hamuy \etal\ 1994). From the published spectro-photometry
for this star and the known central wavelengths and bandwidths,
we estimated the relative fluxes of CD-32\,9927 through our B1 and B2
bandpasses. The intrinsic ratio of the B1/B2 flux for CD-32\,9927 is
1.00871, whereas we observe 1.1036$\pm$0.0010, which implies that
the total throughput of the B1 passband is 9.4\% higher than that
of the B2 passband. This is almost exactly what would be expected
from the throughput of the I$_6$ blocking filter at these wavelengths.

\subsection{Photometry}
\label{rtc}

Photometry was performed using the DAOPHOT package within the IRAF
environment. Experiments were performed with both aperture and point-spread-function (PSF)
photometry. Owing to the variation in seeing observed throughout
a given 30 minute shuffle sequence, it was found that if PSF fitting was to
be used, the PSF had to be evaluated independently for each 1000$\times$90 pixel
sub-segment. This usually resulted in the PSF being determined by just
two stars. However, since our fields are uncrowded, the major reason for
using PSF photometry is to make our photometry more robust against cosmic
ray hits. In this case, the PSF is only being used as a weighting
function, making the small number of stars used to determine it of 
minor significance.

The procedure used was as follows; (1) stars of interest were selected by hand
from a single image; (2) the aperture photometry function of DAOPHOT was then used to
centroid and photometer these stars in all images; (3) PSFs were calculated for
each sub-segment and (4) PSF photometry was obtained using these PSFs. Comparison
of PSF and aperture photometry showed them to produce essentially identical
performance, with the exception that the PSF photometry proved more robust
at rejecting contamination by cosmic rays. It is therefore this PSF data
which we report here.

In order to correct the photometry of our target object for changes in
atmospheric transparency, we adopted the brightest available stars in
each field as zero-point {\em reference} objects. Figure \ref{refseries}a shows the values
of these zero-points for two sample sequences of data -- one taken in photometric
conditions, and the other taken in non-photometric conditions. In order to test these zero-points, 
we further identified a {\em comparison} object in each field, which was not used in
deriving the zero-point. The differential photometry for these {\em comparison} objects is
shown in Fig. \ref{refseries}b for the data shown in Fig. \ref{refseries}a, confirming that
indeed we do remove atmospheric effects. For DENIS-P\,J1228-1547 the {\em comparison}
adopted is 1.3 magnitudes brighter than our target star, while for LP\,944-20 it
is 1.6 magnitudes fainter.

\begin{figure*}
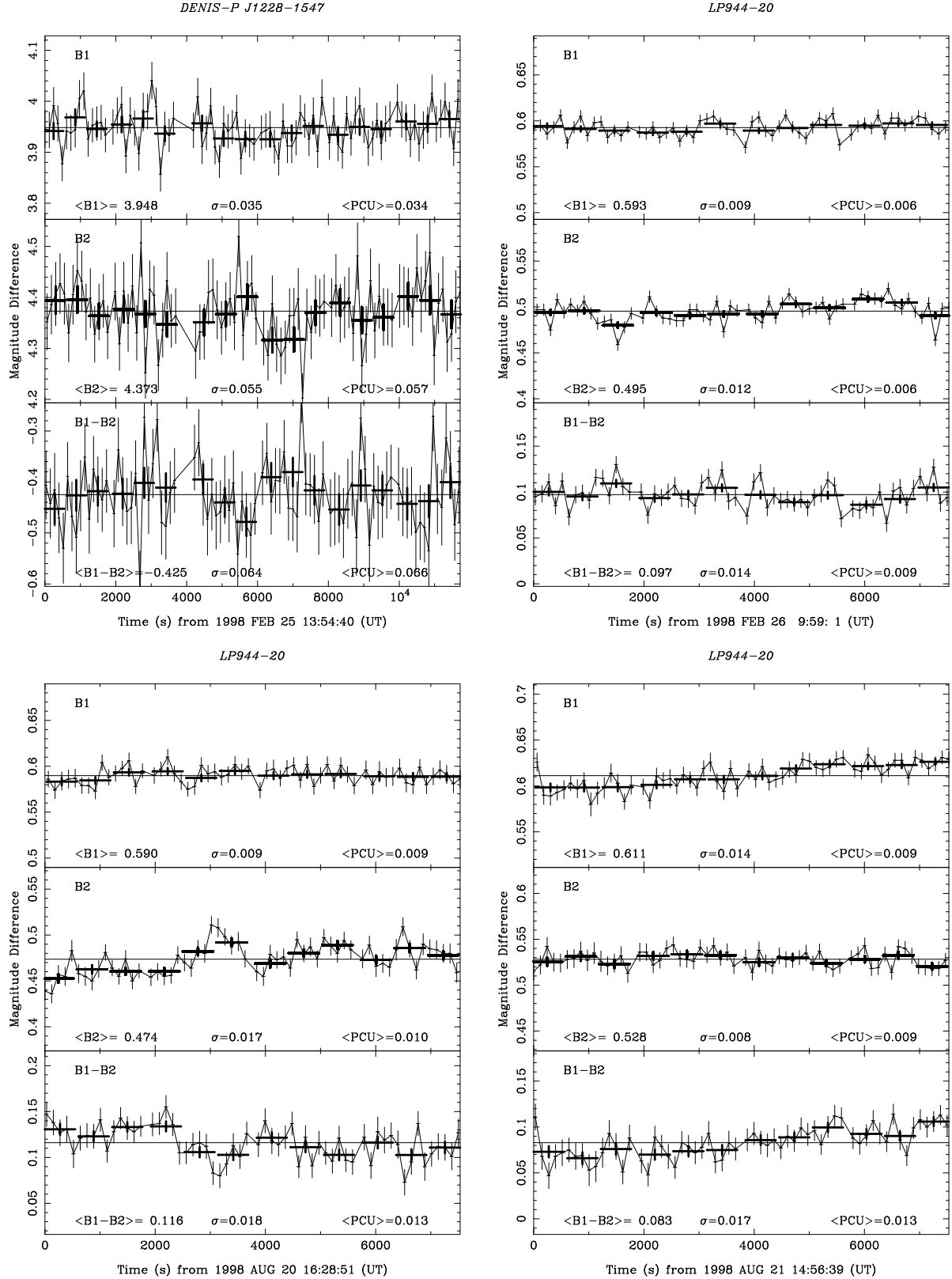

 \centerline{
         \psfig{file=my808f5tl.ps,width=3in}\qquad
         \psfig{file=my808f5tr.ps,width=3in}
            }
 \quad\\
 \centerline{
         \psfig{file=my808f5bl.ps,width=3in}\qquad
         \psfig{file=my808f5br.ps,width=3in}
            }
\caption{Time series data for the observations of the target objects listed in Table \ref{log}. 
In each panel the weighted mean magnitude ($<$B1$>$,$<$B2$>$ and $<$B1--B2$>$), the standard deviation ($\sigma$)
and the mean photon-couting uncertainty ($<$PCU$>$) is shown. We also show as heavy crosses, the data binned as ten minute 
averages (ie 5$\times$1 minute exposures).}
 \label{tseries}
\end{figure*}

Lastly, Fig. \ref{tseries} shows the 
differential photometry for our target objects. In each figure we also show the
colour B1-B2, derived from each B1/B2 observation pair, as well as the
data binned into 10 minute intervals (ie. 5 minutes of exposure time at a duty cycle
of 50\%). The uncertainties plotted
are those produced by the DAOPHOT aperture photometry code (based on photon-counting)
and have been appropriately propagated through all steps. We have plotted the
three observations made of LP\,944-20 separately in each case. Because
these observations are so widely separated in time, we do not consider the
differences in magnitude level {\em between} these observations anywhere near
as significant as differences {\em within} a shuffle sequence.

\begin{figure*}
{\leftline{\large\bf (a)}
 \centerline{
         \psfig{file=my808f4al.ps,width=3in}\qquad
         \psfig{file=my808f4ar.ps,width=3in}
            }
\leftline{\large\bf (b)}
 \centerline{
         \psfig{file=my808f4bl.ps,width=3in}\qquad
         \psfig{file=my808f4br.ps,width=3in}
            }}
\caption{Sample time series data for (a) reference and (b) comparison stars. In each panel
the weighted mean magnitude ($<$B1$>$,$<$B2$>$ and $<$B1--B2$>$), the standard deviation ($\sigma$)
and the mean photon-couting uncertainty ($<$PCU$>$) is shown. In the (b)
panels we also show as heavy crosses, the data binned as ten minute averages
(ie 5$\times$1 minute exposures).}
 \label{refseries}
\end{figure*}

\subsection{Comparison Objects\label{comp}}

In order to determine whether we have made a  significant detection
of variability, it is necessary to define our null hyposthesis -- ie.
what does a non-detection look like.  Kolmogorov-Smirnov (K-S) tests 
(eg. Press \etal\ 1986) were used to test the hypothesis (Hypothesis A)
that the {\em comparison} objects' residual magnitude distributions
were consistent with being drawn from a sample with constant magnitude
and scatter produced by the mean photon-counting uncertainty for that
observing sequence. The results
of these tests are shown in Table \ref{ks}. We conclude from
these that  photon-counting is indeed a reasonable way of estimating the
scatter produced about the mean magnitude of a non-varying object.
This suggests both that our reference objects are not variable, and that
atmospheric effects have been removed to a level where they are smaller
than photon-counting uncertainties.

\subsection{Target Objects\label{targ}}

We can therefore use the photon-counting uncertainties 
for out target objects to
define a null hypothesis for the detection of variability (ie magnitude constancy with scatter
consistent with photon-counting uncertainties -- Hypothesis B) for our target objects.
The results of K-S tests for such a hypothesis are shown in 
Table \ref{ks}.


\begin{table}
  \center
  \caption{Results of Statistical Tests}
  \label{ks}
  \begin{tabular}{lcccccc}
Object$^a$     &  \multicolumn{3}{c}{Hypothesis A$^b$}& \multicolumn{3}{c}{Hypothesis B$^c$}\\
               &   B1     & B2   & B1-B2    & B1     & B2    & B1-B2 \\
               &   (\%)   &(\%)  &(\%)      &(\%)    &(\%)   &(\%)   \\[2pt]
DENIS\,1228         &   59.9   & 56.8   & 38.9   & 99.7   & 62.3   & 28.5   \\
LP\,944-20          &   88.9   & 13.7   & 60.3   & 15.3   & 2.0   & 20.4   \\
LP\,944-20          &   24.2   & 78.0   & 84.7   & 85.7   & 1.1   & 30.4   \\
LP\,944-20          &   94.2   & 68.9   & 92.3   & 2.2   & 97.6  & 38.5  \\
   \end{tabular}
\raggedright
\noindent
\vskip 10pt
$a$ -- the listed order is the same as used in Table 1. \\
$b$ -- see section \ref{comp}. $b$ -- see section \ref{targ}.\\
\end{table}

\section{Discussion}

Due to the shortness of the time series we were able to acquire, it is impossible
to seek any periodicities in this data. We therefore limit ourselves
to setting limits on the extent of variability in these brown dwarfs, and
addressing the cloud implications of this variability.

\subsection{DENIS-P\,J1228-1547}

\begin{figure}
 \centerline{\psfig{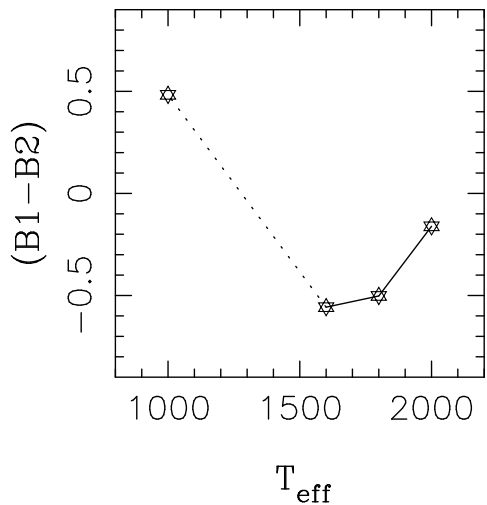}}
 \caption{Calibration curve for differential temperatures as a function of
changes in B1-B2 colours. B1-B2 estimated from spectra of Fig. \ref{spectra}
and \teff\ estimates from the references in Fig. \ref{spectra}. 
Owing to the lack of data in the range {\teff}=1000-1600K,
this portion of the curve is drawn as dashed.
}
 \label{cal}
\end{figure}

Figure \ref{tseries} and Table \ref{ks} fairly convincingly demonstrate that we detect
no evidence for variability in either the B1 or B2 magnitudes, or the B1-B2 colour, in
DENIS-P\,J1228-1547. We therefore ask, what limits this places on the possible presence
of regions of differential \teff\ during our 3 hour observation? The data shown in 
Fig. \ref{tseries} indicate we can
rule out the presence of changes in B1-B2 of greater than 0.096 magnitude over ten minute
periods (ie five 5 minute observations) at the 3-$\sigma$ level. In order to estimate
the \teff\ change such a limit corresponds to, we have integrated the spectra shown in
Fig. \ref{spectra} over our B1 and B2 bandpasses, to obtain a calibration between B1-B2
colour and effective temperature, which we show in Fig. \ref{cal}. We emphasise that 
we aim to use this calibration {\em only}
to estimate temperature changes -- it is {\em not} useful for estimating absolute 
effective temperatures. Unfortunately, no spectra are available in the
\teff=1000--1600K range, though the available data suggest that somewhere in
this range the B1-B2 colour ``turns around'' as the CrH and FeH bands in
which the B2 band lies become weak at very low temperatures (Kirkpatrick \etal\ 1998;
Burrows \& Sharp 1998). The data available, suggest that a B1-B2 colour change of
0.096 mag could correspond to a $\Delta$\teff\ of either $+$220K or $-$70K over the entire
visible disc of the brown dwarf -- with the change to lower temperature being
particularly poorly constrained by the available data. More extreme changes over 
smaller regions of the visible disc are, of course, allowed by these limits.

These limits are not particularly tight, largely because of the weak limits
we place on B1-B2 variation due to the faintness of DENIS-P\,J1228-1547. It is
planned to acquire better data in the future over longer observing
periods in order to improve these limits. There is
also a systematic uncertainty present in the interpretation of these results
due to the lack of objects with published spectra cooler than 1600K. This can
be expected to change as new L-dwarfs are discovered by the 2MASS and DENIS
surveys in the near future (Skrutskie \etal\ 1997; Epchtein \etal\ 1997).

\subsection{LP\,944-20}

The fact that LP\,944-20 is considerably brighter than DENIS-P\,J1228-1547 has
allowed us to acquire much better data for it. The results of the K-S tests
shown in Table \ref{ks} clearly show that in each of the data runs acquired
at least one of the B1 or B2 time series is inconsistent at the 98\% confidence
level with the absence of variability. This can clearly be seen in Figure \ref{tseries}
in both the unbinned and binned data. In particular, we see a rising trend in the
%
%
B2 data for August 20 and in the B1 data for August 21, and short period anomalies
in the February 26 and August 20 B2 data. These effects tend to be washed out in
the colour data by the scatter in the other passband, but remain visible.

\begin{figure}
 \centerline{\psfig{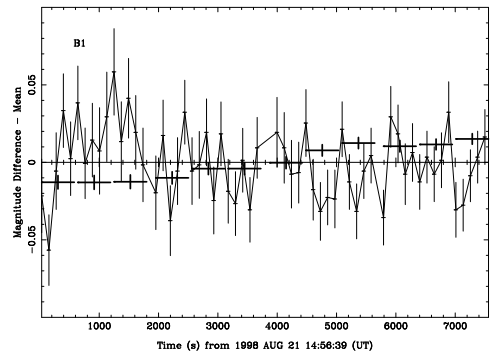}}
 \caption{Comparison of a set of LP\,944-20 binned {\em target} star and unbinned
{\em comparison} star observations. This demonstrates that although the photon-counting
uncertainties for the {\em comparison} star are larger than those LP\,944-20, 
they do not hide the observed variation seen in LP\,944-20 -- the variation seen
is therefore not due to the {\em reference} star adopted.
}
 \label{test}
\end{figure}

A suspicion may remain that the variation we see is
due to the single {\em reference} object adopted, rather than LP\,944-20 itself. 
In Figure \ref{test}
we therefore show the B1 {\em comparison} object data for August 21, with the binned
target object data overlayed. It can be clearly seen that the variation we observe
in LP\,944-20 is not present in the {\em comparison} star, demonstrating that
the variability we see {\em is} due to LP\,944-20.

The data acquired on August 20 and 21 shows tantalising evidence for rotational modulation,
with the B1-B2 colour falling for 2 hours on August 20, and rising on August 21. However,
in light of the fact that we have only 2 hour chunks of data for an object expected
to have a $\approx$4.5 hour period, we make no attempt to set limits on the period.
The fact that variability is so easily detected, however, offers exciting prospects
for period determination from future observations in better conditions and with 
longer observing runs.

We believe the data presented in Fig. \ref{tseries} shows convincing 
evidence for variability in the B1-B2 colour of LP\,944-20
at the 0.04 magnitude (98\% confidence or 2.3-$\sigma$) level. 
Using the calibration shown in Fig. \ref{cal}, this corresponds
to an equivalent $\Delta$\teff\ change of $\approx$20K over the entire
visible disc of the brown dwarf, or $\Delta$\teff$\approx$80K over
25\% of the disc, or $\Delta$\teff$\approx$400K over 5\% of the disc.

The short timescale variation seen in B2 and B1-B2 on August 20 at $\approx$3000\,s
is interesting. With an expected period of $\sim$4.5 hours (if
observed equatorially), the observation of a $\sim$20 minute event would seem to 
imply either that the brown dwarf must have a period significantly shorter
than 4.5 hours (and therefore a significant inclination angle) or that its
photosphere contains a large number of clouds, and that we are observing
the integrated effect of their appearance and disappearance beyond the brown
dwarf's limb. In either case, they offer an exciting prospect for
longer observations and period determination.

It is also interesting that in LP\,944-20, we see evidence for variability in
{\em both} the B1 and B2 magnitudes. Examination of Fig. \ref{spectra} might suggest that
the passage of dust clouds across the disc would be most likely to
be seen in the B1 passband alone, since it is here that the effect of TiO depletion
is strongest. However, the exact height at which dust clouds would form in
L-dwarfs is far from clear  -- the case of Gl\,229B, at least, suggests that some
dust clouds may also form above the photosphere (Burrows \& Sharp 1998), which dim
both B2 and B1, and cause a reddening of B1-B2. As, moreover, is the materials from which 
clouds would form  -- clouds  rich in Cr or Fe, would deplete CrH and FeH 
causing changes in B2 rather than B1).
In other words, the simplest picture of cloud formation is probably an
over-simplification. Further work on brown dwarf meteorology will be
required before we can develop a detailed understanding.

\section{Conclusions: The Future}

We have carried out a first series of observations aimed at observing
meteorological effects in brown dwarfs, using an innovative
charge-shuffling and frequency-switching technique. We find no
evidence for variability in our chosen spectroscopic signature
in the brown dwarf DENIS-P\,J1228-1547, but some evidence for
variability in the hotter brown dwarf LP\,944-20.
The most obvious future development required in
the study of brown dwarf weather is to target a larger sample of
objects with more extensive observations. Several technical
advances also offer exciting prospects. At present we observe each 
passband in 60\,s interleaved exposures. In future, we will implement
a more elegant charge-shuffling scheme to switch between bands
on much shorter ($\sim 10$\,s) timescales, allowing us to
obtain much longer observations in the B1-B2 colour, while
retaining near-simultaneity, and
therefore placing better limits on faint targets. TTF can also be operated at
larger band-widths than were adopted in this experiment,
again improving signal-to-noise on faint targets. And lastly,
we hope to implement frequency-switching in combination with
continuous read-out of our CCDs, allowing us to obtain
arbitrary length time-series, instead of being limited to
the current 30 observations. This improvement will also
allow us to use larger field masks, permitting a larger
ensemble of reference stars. In summary, while the data obtained 
to date shows there
is already reason to believe current techniques can make
headway in the study of brown dwarf weather, future
improvements can be expected to considerably expand our
possibilities.

\subsection*{Acknowledgments}

This paper is based on observations made at the 
Anglo-Australian Telescope, Siding Spring, Australia.
The authors would like to thank M.Marley for helpful comments,
E.Mart\'{\i}n for a thorough referees report, the AAT staff for
their thorough and professional assistance throughout this
observing program, and to thank the TTF team for providing
an instrument of outstanding power and versatility.

\label{lastpage}

\end{document}

%% file: psfig2.tex
\def\PsfigVersion{1.9}
\ifx\undefined\psfig\else \fi

\let\LaTeXAtSign=\@
\let\@=\relax
\edef\psfigRestoreAt{\catcode`\@=\number\catcode`@\relax}
\catcode`\@=11\relax
\newwrite\@unused
\def\ps@typeout#1{{\let\protect\string\immediate\write\@unused{#1}}}
\ps@typeout{psfig/tex \PsfigVersion}

\def\figurepath{./}

\def\@nnil{\@nil}
\def\@empty{}
\def\@psdonoop#1\@@#2#3{}
\def\@psdo#1:=#2\do#3{\edef\@psdotmp{#2}\ifx\@psdotmp\@empty \else
    \expandafter\@psdoloop#2,\@nil,\@nil\@@#1{#3}\fi}
\def\@psdoloop#1,#2,#3\@@#4#5{\def#4{#1}\ifx #4\@nnil \else
       #5\def#4{#2}\ifx #4\@nnil \else#5\@ipsdoloop #3\@@#4{#5}\fi\fi}
\def\@ipsdoloop#1,#2\@@#3#4{\def#3{#1}\ifx #3\@nnil 
       \let\@nextwhile=\@psdonoop \else
      #4\relax\let\@nextwhile=\@ipsdoloop\fi\@nextwhile#2\@@#3{#4}}
\def\@tpsdo#1:=#2\do#3{\xdef\@psdotmp{#2}\ifx\@psdotmp\@empty \else
    \@tpsdoloop#2\@nil\@nil\@@#1{#3}\fi}
\def\@tpsdoloop#1#2\@@#3#4{\def#3{#1}\ifx #3\@nnil 
       \let\@nextwhile=\@psdonoop \else
      #4\relax\let\@nextwhile=\@tpsdoloop\fi\@nextwhile#2\@@#3{#4}}
\ifx\undefined\fbox
\newdimen\fboxrule
\newdimen\fboxsep
\newdimen\ps@tempdima
\newbox\ps@tempboxa
\long\def\fbox#1{\leavevmode\setbox\ps@tempboxa\hbox{#1}\ps@tempdima\fboxrule
    \advance\ps@tempdima \fboxsep \advance\ps@tempdima \dp\ps@tempboxa
   \hbox{\lower \ps@tempdima\hbox
  {\vbox{\hrule height \fboxrule
          \hbox{\vrule width \fboxrule \hskip\fboxsep
          \vbox{\vskip\fboxsep \box\ps@tempboxa\vskip\fboxsep}\hskip 
                 \fboxsep\vrule width \fboxrule}
                 \hrule height \fboxrule}}}}
\fi
\newread\ps@stream
\newif\ifnot@eof       
\newif\if@noisy        
\newif\if@atend        
\newif\if@psfile       
{\catcode`\%=12\global\gdef\epsf@start{
\def\epsf@PS{PS}
\def\epsf@getbb#1{%
\openin\ps@stream=#1
\ifeof\ps@stream\ps@typeout{Error, File #1 not found}\else
   {\not@eoftrue \chardef\other=12
    \def\do##1{\catcode`##1=\other}\dospecials \catcode`\ =10
    \loop
       \if@psfile
	  \read\ps@stream to \epsf@fileline
       \else{
	  \obeyspaces
          \read\ps@stream to \epsf@tmp\global\let\epsf@fileline\epsf@tmp}
       \fi
       \ifeof\ps@stream\not@eoffalse\else
       \if@psfile\else
       \expandafter\epsf@test\epsf@fileline:. \\%
       \fi
          \expandafter\epsf@aux\epsf@fileline:. \\%
       \fi
   \ifnot@eof\repeat
   }\closein\ps@stream\fi}%
\long\def\epsf@test#1#2#3:#4\\{\def\epsf@testit{#1#2}
			\ifx\epsf@testit\epsf@start\else
\ps@typeout{Warning! File does not start with `\epsf@start'.  It may not be a PostScript file.}
			\fi
			\@psfiletrue} 
{\catcode`\%=12\global\let\epsf@percent=
\long\def\epsf@aux#1#2:#3\\{\ifx#1\epsf@percent
   \def\epsf@testit{#2}\ifx\epsf@testit\epsf@bblit
	\@atendfalse
        \epsf@atend #3 . \\%
	\if@atend	
	   \if@verbose{
		\ps@typeout{psfig: found `(atend)'; continuing search}
	   }\fi
        \else
        \epsf@grab #3 . . . \\%
        \not@eoffalse
        \global\no@bbfalse
        \fi
   \fi\fi}%
\def\epsf@grab #1 #2 #3 #4 #5\\{%
   \global\def\epsf@llx{#1}\ifx\epsf@llx\empty
      \epsf@grab #2 #3 #4 #5 .\\\else
   \global\def\epsf@lly{#2}%
   \global\def\epsf@urx{#3}\global\def\epsf@ury{#4}\fi}%
\def\epsf@atendlit{(atend)} 
\def\epsf@atend #1 #2 #3\\{%
   \def\epsf@tmp{#1}\ifx\epsf@tmp\empty
      \epsf@atend #2 #3 .\\\else
   \ifx\epsf@tmp\epsf@atendlit\@atendtrue\fi\fi}

\chardef\psletter = 11 
\chardef\other = 12

\newif \ifdebug 
\newif\ifc@mpute 
\c@mputetrue 

\let\then = \relax
\def\r@dian{pt }
\let\r@dians = \r@dian
\let\dimensionless@nit = \r@dian
\let\dimensionless@nits = \dimensionless@nit
\def\internal@nit{sp }
\let\internal@nits = \internal@nit
\newif\ifstillc@nverging
\def \Mess@ge #1{\ifdebug \then \message {#1} \fi}

{ 
	\catcode `\@ = \psletter
	\gdef \nodimen {\expandafter \n@dimen \the \dimen}
	\gdef \term #1 #2 #3%
	       {\edef \t@ {\the #1}
		\edef \t@@ {\expandafter \n@dimen \the #2\r@dian}%
		\t@rm {\t@} {\t@@} {#3}%
	       }
	\gdef \t@rm #1 #2 #3%
	       {{%
		\count 0 = 0
		\dimen 0 = 1 \dimensionless@nit
		\dimen 2 = #2\relax
		\Mess@ge {Calculating term #1 of \nodimen 2}%
		\loop
		\ifnum	\count 0 < #1
		\then	\advance \count 0 by 1
			\Mess@ge {Iteration \the \count 0 \space}%
			\Multiply \dimen 0 by {\dimen 2}%
			\Mess@ge {After multiplication, term = \nodimen 0}%
			\Divide \dimen 0 by {\count 0}%
			\Mess@ge {After division, term = \nodimen 0}%
		\repeat
		\Mess@ge {Final value for term #1 of 
				\nodimen 2 \space is \nodimen 0}%
		\xdef \Term {#3 = \nodimen 0 \r@dians}%
		\aftergroup \Term
	       }}
	\catcode `\p = \other
	\catcode `\t = \other
	\gdef \n@dimen #1pt{#1} 
}

\def \Divide #1by #2{\divide #1 by #2} 

\def \Multiply #1by #2
       {{
	\count 0 = #1\relax
	\count 2 = #2\relax
	\count 4 = 65536
	\Mess@ge {Before scaling, count 0 = \the \count 0 \space and
			count 2 = \the \count 2}%
	\ifnum	\count 0 > 32767 
	\then	\divide \count 0 by 4
		\divide \count 4 by 4
	\else	\ifnum	\count 0 < -32767
		\then	\divide \count 0 by 4
			\divide \count 4 by 4
		\else
		\fi
	\fi
	\ifnum	\count 2 > 32767 
	\then	\divide \count 2 by 4
		\divide \count 4 by 4
	\else	\ifnum	\count 2 < -32767
		\then	\divide \count 2 by 4
			\divide \count 4 by 4
		\else
		\fi
	\fi
	\multiply \count 0 by \count 2
	\divide \count 0 by \count 4
	\xdef \product {#1 = \the \count 0 \internal@nits}%
	\aftergroup \product
       }}

\def\r@duce{\ifdim\dimen0 > 90\r@dian \then   
		\multiply\dimen0 by -1
		\advance\dimen0 by 180\r@dian
		\r@duce
	    \else \ifdim\dimen0 < -90\r@dian \then  
		\advance\dimen0 by 360\r@dian
		\r@duce
		\fi
	    \fi}

\def\Sine#1%
       {{%
	\dimen 0 = #1 \r@dian
	\r@duce
	\ifdim\dimen0 = -90\r@dian \then
	   \dimen4 = -1\r@dian
	   \c@mputefalse
	\fi
	\ifdim\dimen0 = 90\r@dian \then
	   \dimen4 = 1\r@dian
	   \c@mputefalse
	\fi
	\ifdim\dimen0 = 0\r@dian \then
	   \dimen4 = 0\r@dian
	   \c@mputefalse
	\fi
	\ifc@mpute \then
		\divide\dimen0 by 180
		\dimen0=3.141592654\dimen0
		\dimen 2 = 3.1415926535897963\r@dian 
		\divide\dimen 2 by 2 
		\Mess@ge {Sin: calculating Sin of \nodimen 0}%
		\count 0 = 1 
		\dimen 2 = 1 \r@dian 
		\dimen 4 = 0 \r@dian 
		\loop
			\ifnum	\dimen 2 = 0 
			\then	\stillc@nvergingfalse 
			\else	\stillc@nvergingtrue
			\fi
			\ifstillc@nverging 
			\then	\term {\count 0} {\dimen 0} {\dimen 2}%
				\advance \count 0 by 2
				\count 2 = \count 0
				\divide \count 2 by 2
				\ifodd	\count 2 
				\then	\advance \dimen 4 by \dimen 2
				\else	\advance \dimen 4 by -\dimen 2
				\fi
		\repeat
	\fi		
			\xdef \sine {\nodimen 4}%
       }}

\def\Cosine#1{\ifx\sine\UnDefined\edef\Savesine{\relax}\else
		             \edef\Savesine{\sine}\fi
	{\dimen0=#1\r@dian\advance\dimen0 by 90\r@dian
	 \Sine{\nodimen 0}
	 \xdef\cosine{\sine}
	 \xdef\sine{\Savesine}}}	      

\def\psdraft{
	\def\@psdraft{0}
}
\def\psfull{
	\def\@psdraft{100}
}

\psfull

\newif\if@scalefirst
\def\psscalefirst{\@scalefirsttrue}
\def\psrotatefirst{\@scalefirstfalse}
\psrotatefirst

\newif\if@draftbox
\def\psnodraftbox{
	\@draftboxfalse
}
\def\psdraftbox{
	\@draftboxtrue
}
\@draftboxtrue

\newif\if@prologfile
\newif\if@postlogfile
\def\pssilent{
	\@noisyfalse
}
\def\psnoisy{
	\@noisytrue
}
\psnoisy
\newif\if@bbllx
\newif\if@bblly
\newif\if@bburx
\newif\if@bbury
\newif\if@height
\newif\if@width
\newif\if@rheight
\newif\if@rwidth
\newif\if@angle
\newif\if@clip
\newif\if@verbose
\newif\if@scale
\def\@p@@sclip#1{\@cliptrue}

\newif\if@decmpr

\def\@p@@sfigure#1{\def\@p@sfile{null}\def\@p@sbbfile{null}
	        \openin1=#1.bb
		\ifeof1\closein1
	        	\openin1=\figurepath#1.bb
			\ifeof1\closein1
			        \openin1=#1
				\ifeof1\closein1%
				       \openin1=\figurepath#1
					\ifeof1
					   \ps@typeout{Error, File #1 not found}
						\if@bbllx\if@bblly
				   		\if@bburx\if@bbury
			      				\def\@p@sfile{#1}%
			      				\def\@p@sbbfile{#1}%
							\@decmprfalse
				  	   	\fi\fi\fi\fi
					\else\closein1
				    		\def\@p@sfile{\figurepath#1}%
				    		\def\@p@sbbfile{\figurepath#1}%
						\@decmprfalse
	                       		\fi%
			 	\else\closein1%
					\def\@p@sfile{#1}
					\def\@p@sbbfile{#1}
					\@decmprfalse
			 	\fi
			\else
				\def\@p@sfile{\figurepath#1}
				\def\@p@sbbfile{\figurepath#1.bb}
				\@decmprtrue
			\fi
		\else
			\def\@p@sfile{#1}
			\def\@p@sbbfile{#1.bb}
			\@decmprtrue
		\fi}

\def\@p@@sfile#1{\@p@@sfigure{#1}}

\def\@p@@sbbllx#1{
		\@bbllxtrue
		\dimen100=#1
		\edef\@p@sbbllx{\number\dimen100}
}
\def\@p@@sbblly#1{
		\@bbllytrue
		\dimen100=#1
		\edef\@p@sbblly{\number\dimen100}
}
\def\@p@@sbburx#1{
		\@bburxtrue
		\dimen100=#1
		\edef\@p@sbburx{\number\dimen100}
}
\def\@p@@sbbury#1{
		\@bburytrue
		\dimen100=#1
		\edef\@p@sbbury{\number\dimen100}
}
\def\@p@@sheight#1{
		\@heighttrue
		\dimen100=#1
   		\edef\@p@sheight{\number\dimen100}
}
\def\@p@@swidth#1{
		\@widthtrue
		\dimen100=#1
		\edef\@p@swidth{\number\dimen100}
}
\def\@p@@srheight#1{
		\@rheighttrue
		\dimen100=#1
		\edef\@p@srheight{\number\dimen100}
}
\def\@p@@srwidth#1{
		\@rwidthtrue
		\dimen100=#1
		\edef\@p@srwidth{\number\dimen100}
}
\def\@p@@sangle#1{
		\@angletrue
		\edef\@p@sangle{#1} 
}
\def\@p@@srotate#1{\@p@@sangle{-#1}}
\def\@p@@sscale#1{
		\@scaletrue
		\edef\@p@sscale{#1}
}
\def\@p@@ssilent#1{ 
		\@verbosefalse
}
\def\@p@@sprolog#1{\@prologfiletrue\def\@prologfileval{#1}}
\def\@p@@spostlog#1{\@postlogfiletrue\def\@postlogfileval{#1}}
\def\@cs@name#1{\csname #1\endcsname}
\def\@setparms#1=#2,{\@cs@name{@p@@s#1}{#2}}
\def\ps@init@parms{
		\@bbllxfalse \@bbllyfalse
		\@bburxfalse \@bburyfalse
		\@heightfalse \@widthfalse
		\@rheightfalse \@rwidthfalse
		\@scalefalse
		\def\@p@sbbllx{}\def\@p@sbblly{}
		\def\@p@sbburx{}\def\@p@sbbury{}
		\def\@p@sheight{}\def\@p@swidth{}
		\def\@p@srheight{}\def\@p@srwidth{}
		\def\@p@sangle{0}
		\def\@p@sfile{} \def\@p@sbbfile{}
		\def\@p@scost{10}
		\def\@sc{}
		\@prologfilefalse
		\@postlogfilefalse
		\@clipfalse
		\if@noisy
			\@verbosetrue
		\else
			\@verbosefalse
		\fi
}
\def\parse@ps@parms#1{
	 	\@psdo\@psfiga:=#1\do
		   {\expandafter\@setparms\@psfiga,}}
\newif\ifno@bb
\def\bb@missing{
	\if@verbose{
		\ps@typeout{psfig: searching \@p@sbbfile \space  for bounding box}
	}\fi
	\no@bbtrue
	\epsf@getbb{\@p@sbbfile}
        \ifno@bb \else \bb@cull\epsf@llx\epsf@lly\epsf@urx\epsf@ury\fi
}	
\def\bb@cull#1#2#3#4{
	\dimen100=#1 bp\edef\@p@sbbllx{\number\dimen100}
	\dimen100=#2 bp\edef\@p@sbblly{\number\dimen100}
	\dimen100=#3 bp\edef\@p@sbburx{\number\dimen100}
	\dimen100=#4 bp\edef\@p@sbbury{\number\dimen100}
	\no@bbfalse
}
\newdimen\p@intvaluex
\newdimen\p@intvaluey
\def\rotate@#1#2{{\dimen0=#1 sp\dimen1=#2 sp
		  \global\p@intvaluex=\cosine\dimen0
		  \dimen3=\sine\dimen1
		  \global\advance\p@intvaluex by -\dimen3
		  \global\p@intvaluey=\sine\dimen0
		  \dimen3=\cosine\dimen1
		  \global\advance\p@intvaluey by \dimen3
		  }}
\def\compute@bb{
		\no@bbfalse
		\if@bbllx \else \no@bbtrue \fi
		\if@bblly \else \no@bbtrue \fi
		\if@bburx \else \no@bbtrue \fi
		\if@bbury \else \no@bbtrue \fi
		\ifno@bb \bb@missing \fi
		\ifno@bb \ps@typeout{FATAL ERROR: no bb supplied or found}
			\no-bb-error
		\fi
		\count203=\@p@sbburx
		\count204=\@p@sbbury
		\advance\count203 by -\@p@sbbllx
		\advance\count204 by -\@p@sbblly
		\edef\ps@bbw{\number\count203}
		\edef\ps@bbh{\number\count204}
		\if@angle 
			\Sine{\@p@sangle}\Cosine{\@p@sangle}
	        	{\dimen100=\maxdimen\xdef\r@p@sbbllx{\number\dimen100}
					    \xdef\r@p@sbblly{\number\dimen100}
			                    \xdef\r@p@sbburx{-\number\dimen100}
					    \xdef\r@p@sbbury{-\number\dimen100}}
                        \def\minmaxtest{
			   \ifnum\number\p@intvaluex<\r@p@sbbllx
			      \xdef\r@p@sbbllx{\number\p@intvaluex}\fi
			   \ifnum\number\p@intvaluex>\r@p@sbburx
			      \xdef\r@p@sbburx{\number\p@intvaluex}\fi
			   \ifnum\number\p@intvaluey<\r@p@sbblly
			      \xdef\r@p@sbblly{\number\p@intvaluey}\fi
			   \ifnum\number\p@intvaluey>\r@p@sbbury
			      \xdef\r@p@sbbury{\number\p@intvaluey}\fi
			   }
			\rotate@{\@p@sbbllx}{\@p@sbblly}
			\minmaxtest
			\rotate@{\@p@sbbllx}{\@p@sbbury}
			\minmaxtest
			\rotate@{\@p@sbburx}{\@p@sbblly}
			\minmaxtest
			\rotate@{\@p@sbburx}{\@p@sbbury}
			\minmaxtest
			\edef\@p@sbbllx{\r@p@sbbllx}\edef\@p@sbblly{\r@p@sbblly}
			\edef\@p@sbburx{\r@p@sbburx}\edef\@p@sbbury{\r@p@sbbury}
		\fi
		\count203=\@p@sbburx
		\count204=\@p@sbbury
		\advance\count203 by -\@p@sbbllx
		\advance\count204 by -\@p@sbblly
		\edef\@bbw{\number\count203}
		\edef\@bbh{\number\count204}
}
\def\in@hundreds#1#2#3{\count240=#2 \count241=#3
		     \count100=\count240	
		     \divide\count100 by \count241
		     \count101=\count100
		     \multiply\count101 by \count241
		     \advance\count240 by -\count101
		     \multiply\count240 by 10
		     \count101=\count240	
		     \divide\count101 by \count241
		     \count102=\count101
		     \multiply\count102 by \count241
		     \advance\count240 by -\count102
		     \multiply\count240 by 10
		     \count102=\count240	
		     \divide\count102 by \count241
		     \count200=#1\count205=0
		     \count201=\count200
			\multiply\count201 by \count100
		 	\advance\count205 by \count201
		     \count201=\count200
			\divide\count201 by 10
			\multiply\count201 by \count101
			\advance\count205 by \count201
		     \count201=\count200
			\divide\count201 by 100
			\multiply\count201 by \count102
			\advance\count205 by \count201
		     \edef\@result{\number\count205}
}
\def\ps@scaleinhundreds#1{
		\in@hundreds{#1}{\@p@sscale}{100}
		\edef#1{\@result}
}
\def\compute@wfromh{
		\in@hundreds{\@p@sheight}{\@bbw}{\@bbh}
		\edef\@p@swidth{\@result}
}
\def\compute@hfromw{
	        \in@hundreds{\@p@swidth}{\@bbh}{\@bbw}
		\edef\@p@sheight{\@result}
}
\def\compute@handw{
		\if@height 
			\if@width
			\else
				\compute@wfromh
			\fi
		\else 
			\if@width
				\compute@hfromw
			\else
				\edef\@p@sheight{\@bbh}
				\edef\@p@swidth{\@bbw}
			\fi
		\fi
}
\def\compute@resv{
		\if@rheight \else \edef\@p@srheight{\@p@sheight} \fi
		\if@rwidth \else \edef\@p@srwidth{\@p@swidth} \fi
}
\def\compute@sizes{
	\compute@bb
	\if@scalefirst\if@angle
	\if@width
	   \in@hundreds{\@p@swidth}{\@bbw}{\ps@bbw}
	   \edef\@p@swidth{\@result}
	\fi
	\if@height
	   \in@hundreds{\@p@sheight}{\@bbh}{\ps@bbh}
	   \edef\@p@sheight{\@result}
	\fi
	\fi\fi
	\compute@handw
	\compute@resv
	\if@scale
	   \if@verbose
	      \ps@typeout{(scaling by \@p@sscale)}%
	   \fi
	   \ps@scaleinhundreds{\@p@swidth}%
	   \ps@scaleinhundreds{\@p@sheight}%
	   \ps@scaleinhundreds{\@p@srwidth}%
	   \ps@scaleinhundreds{\@p@srheight}%
	\fi
}

\def\psfig#1{\vbox {
	\ps@init@parms
	\parse@ps@parms{#1}
	\compute@sizes
	\ifnum\@p@scost<\@psdraft{
		\special{ps::[begin] 	\@p@swidth \space \@p@sheight \space
				\@p@sbbllx \space \@p@sbblly \space
				\@p@sbburx \space \@p@sbbury \space
				startTexFig \space }
		\if@angle
			\special {ps:: \@p@sangle \space rotate \space} 
		\fi
		\if@clip{
			\if@verbose{
				\ps@typeout{(clip)}
			}\fi
			\special{ps:: doclip \space }
		}\fi
		\if@prologfile
		    \special{ps: plotfile \@prologfileval \space } \fi
		\if@decmpr{
			\if@verbose{
				\ps@typeout{psfig: including \@p@sfile.Z \space }
			}\fi
			\special{ps: plotfile "`zcat \@p@sfile.Z" \space }
		}\else{
			\if@verbose{
				\ps@typeout{psfig: including \@p@sfile \space }
			}\fi
			\special{ps: plotfile \@p@sfile \space }
		}\fi
		\if@postlogfile
		    \special{ps: plotfile \@postlogfileval \space } \fi
		\special{ps::[end] endTexFig \space }
		\vbox to \@p@srheight true sp{
			\hbox to \@p@srwidth true sp{
				\hss
			}
		\vss
		}
	}\else{
		\if@draftbox{		
			\hbox{\frame{\vbox to \@p@srheight true sp{
			\vss
			\hbox to \@p@srwidth true sp{ \hss \@p@sfile \hss }
			\vss
			}}}
		}\else{
			\vbox to \@p@srheight true sp{
			\vss
			\hbox to \@p@srwidth true sp{\hss}
			\vss
			}
		}\fi

	}\fi
}}
\psfigRestoreAt
\let\@=\LaTeXAtSign